\documentclass[showpacs,pre]{revtex4}
\usepackage{amsmath}
\usepackage{graphicx}
\begin{document}
\title{Aftershocks and Omori's law in a modified Carlson-Langer model with nonlinear visco-elasticity}
\author{Hidetsugu Sakaguchi and Kazuki Okamura}
\affiliation{Department of Applied Science for Electronics and Materials,
Interdisciplinary Graduate School of Engineering Sciences, Kyushu
University, Kasuga, Fukuoka 816-8580, Japan}
\begin{abstract}
A modified Carlson-Langer model for earthquakes is proposed, which includes nonlinear visco-elasticity. Several aftershocks are generated after the main shock owing to the damping of the additional visco-elastic force. Both the Gutenberg-Richter law and Omori's law are reproduced in a numerical simulation of the modified Carlson-Langer model on a critical percolation cluster of a square lattice. 
\end{abstract}
\maketitle
\section{Introduction}
Earthquakes follow various types of power laws. 
Gutenberg and Richter found that  the slip-size of faults or the seismic moment $S$ obeys a power law:~\cite{rf:1} 
\begin{equation}
P(M)\sim 10^{-bM},
\end{equation}
where the magnitude $M$ is defined as $M=(2/3)\log_{10}S$. The exponent $b$ takes a value ranging from 0.8 to 1.2, and the most typical value is $b=1$. 
After a large main shock, many aftershocks occur. Omori found that the frequency of aftershocks decays in a power law:~\cite{rf:2,rf:3} 
\begin{equation}
n(\tau)=\frac{K}{(c+\tau)^p},
\end{equation}
where $\tau$ is the elapsed time from the main shock, and the exponent $p$ is expected to take a value between 0.9  and 1.4.  In the original Omori's law, $p$ is set to be 1. 
Omori's law implies that the frequency of aftershocks decays slowly or 
aftershocks can occur even after a rather long time following the main shock.  
The mechanisms of earthquakes and aftershocks have been studied by many authors, but are not completely understood. 
There are several deterministic models for earthquakes. The power laws of earthquakes might be understood from the chaotic dynamics of the nonlinear equations.  Burridge and Knopoff proposed a block-spring model for earthquakes~\cite{rf:4}.
Later, Carlson and Langer used a type of velocity-weakening friction and found a power law in the small magnitude range~\cite{rf:5}.  
Many authors studied intensively the Carlson-Langer model and its modification. Vieira, Vasconcelos, Nagel pointed out a transition in the magnitude distribution by changing a parameter of the velocity-weakening friction~\cite{rf:6}. 
On the other hand, there are various models in which the power laws originate from the fractal geometry of faults or asperities of plate boundaries. The size distribution of earthquakes has been explained by several authors using such fractal geometry~\cite{rf:7}.  
We studied the block-spring system on critical percolation clusters with fractal geometry, and showed that the exponent of the power law is about $1$, which is close to the exponent in the original Gutenberg-Richter law~\cite{rf:8}.
There are also several stochastic cellular automaton models for earthquakes. For example, Bak and Tang proposed a cellular automaton model for earthquakes which exhibits self-organized criticality~\cite{rf:9}.  Olami, Feder, and Christensen proposed a modified cellular automaton model with some dissipation~\cite{rf:10}.  

The difference between main shocks and aftershocks is not clear in the Carlson-Langer model or the Bak-Tang model. Ito and Matsuzaki proposed a modified model of Bak-Tang model and reproduced Omori's law for aftershocks by randomly perturbing certain regions that slipped during the main shock~\cite{rf:11}. Dieterich studied time-dependent friction in rocks and suggested a relationship between the visco-elasticity and  Omori's law~\cite{rf:12}. Hainzl et al. included some visco-elasticity in the Olami-Feder-Christensen model and reproduced a version of Omori's law~\cite{rf:13}. Jagla et al. studied a depinning model of viscoelastic interfaces and reproduced a type of Omori's law~\cite{rf:14,rf:15}. 
However, the physical meaning of models constructed by Ito-Matsuzaki and Hainzl et al. remains unclear. Jagla's depinning model includes randomness and the relationship between the depinning model and earthquakes is not obvious. In this paper, we study a modified version of the Carlson-Langer model with visco-elastic elements, and discuss Omori's law of aftershocks. Our model is based on mechanics or Newton's equation of motion, and does not include randomness or external noise in contrast to the models by Hainzl et al. and Jagla et al. The time evolution is continuous during slip events, but stick intervals between successive slips are theoretically evaluated and numerical simulation for the stick intervals is skipped. A rather long-time numerical simulation becomes possible owing to the reduction in computation time. 
 
\section{The Carlson-Langer model with viscoelasticity}
The Carlson-Langer model is a block-spring model. 
In a one-dimensional model, $N$ blocks of mass $m=1$ are coupled with the nearest neighbor blocks using a spring with a spring constant $k$. In the original model, each block is further coupled with a spring to a rigid plate moving with a constant velocity $v_0$. The model equation is expressed as 
\begin{equation}
\frac{d^2x_i}{dt^2}=k(x_{i+1}-2x_i+x_{i-1})+(v_0t-x_i)-\phi(v_i),
\end{equation}
where $v_i$ denotes the velocity of the block $dx_i/dt$, $k$ is the spring constant between neighboring blocks, and $\phi(v_i)$ represents the kinetic friction. The maximum static friction is set to be 1. In the Carlson-Langer model, the kinetic friction $\phi(v)$ is assumed to be 
\begin{equation}
\phi(v)=\frac{1-\sigma}{1+2\alpha v/(1-\sigma)}
\end{equation}
for $v>0$. Two parameters $\alpha$ and $\sigma$ characterize the velocity-weakening friction. The parameter $\sigma$ represents the difference between the maximum static friction 1 and the kinetic friction at the limit of $v=0$. The parameters $k$ and $\sigma$ are fixed to be $k=16$ and $\sigma=0.01$ in our numerical simulations for the sake of simplicity.  
The parameter $\alpha$ expresses the velocity-weakening rate of the kinetic friction, i.e., $d\phi(v)/dv=-2\alpha$ at the limit of $v=0$. When $v_i$ decreases to 0, the slip state changes to the stick state.  The stick state is assumed to be maintained until the applied force reaches the maximum static friction 1. It is assumed that the reverse motion does not occur~\cite{rf:5}. 

In the modified model, a dashpot and spring are added in parallel to the spring coupled with the rigid plate pulled by velocity $v_0$ as shown in Fig.~1.   
An additional force $f_i\le 0$ is applied to the $i$th block by the additional dashpot and spring after the block starts to slip. The additional force works as a braking force to the slipping block. 
Various models including viscoelasticity are possible. For example, the dashpot and the spring can be added in parallel to the spring connecting neighboring blocks. These models might be interesting but we investigate only the simplest model shown in Fig.~1 in this paper. One reason is that the role of the viscoelasticity as a brake force is rather clear and anther reason is that the dynamics in the stick phase is simple and analytically calculated as explained in later paragraphs.    

The force satisfies $f_i=Ky_{i}$ where $-y_{i}$ represents the contraction of the additional spring. For a linear dashpot, the force $f_i$ satisfies $f_i=-\eta (-u_i)$ where $u_{i}<0$ denotes the slip velocity at the dashpot. 
Because Omori's law is not reproduced in a model using the linear dashpot as shown later, we use a more general model where $f_i$ satisfies $f_i=-\eta (-u_{i})^{\gamma}$. A power-law relation is assumed between the slip velocity and resistance force, and $\gamma=1$ corresponds to the linear dashpot. 
A power-law model for the relationship between shear velocity and shear stress is one of the typical models for non-Newton fluids. 
The power-law model of $0<\gamma<1$ is applied to non-Newton fluids such as polymeric fluids, tooth paste and chocolate.  The slip velocity $u_i$ is expressed as $u_{i}=-(-f_i/\eta)^{\beta}$, where $\beta=1/\gamma$.   
The velocity difference between the upper plate and the $i$th block is written as $v_0-dx_i/dt$ which is equal to $dy_{i}/dt+u_{i}$. Then, $x_i$ and $f_i$ satisfy
\begin{eqnarray}
\frac{d^2x_i}{dt^2}&=&k(x_{i+1}-2x_i+x_{i-1})+(v_0t-x_i)+f_i-\phi(v_i),\nonumber\\
\frac{df_i}{dt}&=&K\frac{dy_{i}}{dt}=K(v_0-\frac{dx_i}{dt}-u_{i})=K\left (v_0-\frac{dx_i}{dt}\right )+K\left (\frac{-f_i}{\eta}\right )^{\beta}.
\end{eqnarray}
\begin{figure}[t]
\begin{center}
\includegraphics[height=3.cm]{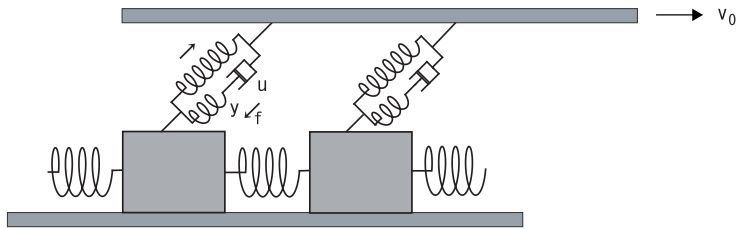}
\end{center}
\caption{Schematic figure of a modified Carlson-Langer model with visco-elasticity.}
\label{f1}
\end{figure}

We further assume that $v_0$ is infinitesimally small, and $v_0$ is set to be zero during the slip phase. In the stationary stick phase, $f_i=0$, and $v_i=0$. 
First, we calculate the maximum value $F$ of $F_i=k(x_{i+1}-2x_i+x_{i-1})+(v_0t-x_i)$ among the $N$ blocks. If the $i$th block takes the maximum value, the time is advanced from $t$ to $t+(1-F)/v_0$. The pulling force at the $i$th block exceeds the maximum static friction $1$, and the $i$th block starts to slip. The neighboring blocks can slip together owing to the elastic coupling of spring constant $k$. The time evolution of the slip process obeys Eq.~(5). 
Because we assume $v_0$ is infinitesimally small, two slip events do not occur simultaneously in distant places. 
When the first slip event (main shock) starts, $dx_i/dt>0$ for the block $i$, and $f_i$ becomes negative.  That is, the braking force $f_i$ works for the slipping block.  After the main shock ends and the stick phase starts,  the slip velocity $dx_i/dt$ becomes  zero, then $f_i<0$ increases toward zero by the last term $K(-f_i/\eta)^{\beta}>0$ owing to the effect of the dashpot. That is, the braking force dampens and  the total pulling force to the block tends to increase over time. It is possible that some blocks start to slip again after a certain time following the main shock. This is an aftershock. Aftershocks can occur repeatedly. After the sequence of aftershocks, all $f_i$'s decay to zero. The period from the start of a main shock and the end of the sequence of aftershocks is one shock event.  At the end of the shock event, $F_i=k(x_{i+1}-2x_i+x_{i-1})+(v_0t-x_i)$ is smaller than 1 and $f_i=0$ for all $i$'s. 
Then, we again calculate the maximum value $F$ of $F_i=k(x_{i+1}-2x_i+x_{i-1})+(v_0t-x_i)$ among the $N$ blocks, and the upper plate is slowly moved until another block begins to slip.  The time interval $\Delta t$ until the starting time of the next main shock is calculated by $v_0\Delta t=1-F$. 
Similarly, we can calculate the start time of aftershocks in the sequence of aftershocks. In the stick phase after an aftershock, $f_i(t)$ can be explicitly expressed as
\begin{equation}
\{-f_i(t)\}^{1-\beta}=\{-f_i(t_0)\}^{1-\beta}+(1-\beta)\frac{K(t-t_0)}{\eta^{\beta}},
\end{equation}
for $\beta\ne 1$ because $v_0=0$ and $dx_i/dt=0$ after the final time $t_0$ of the aftershock. From Eq.~(6), a possible start time $t_i$ of the next aftershock for the $i$th block is evaluated as
\begin{equation}
t_i=t_0+\frac{\eta^{\beta}}{K(1-\beta)}\{(F_i-1)^{1-\beta}-(-f_i(t_0))^{1-\beta}\},
\end{equation}
because $F_i+f_i(t)=1$ at the start time $t_i$ of the next aftershock. 
The minimum value $t_s$ of $t_i$ is the actual start time of the next aftershock, and $f_i(t_s)$ at $t=t_s$ is evaluated using Eq.~(6).  
If all $F_i$'s are smaller than 1, the sequence of aftershocks ends. 
In the case of the linear visco-elasticity: $\gamma=\beta=1$, Eqs.(6) and (7) are replaced by 
\begin{equation}
f_i(t)=f_i(t_0)e^{-K(t-t_0)/\eta},
\end{equation}
and 
\begin{equation}
t_i=t_0-\frac{\eta}{K}\ln\frac{F_i-1}{(-f_i(t_0))}.
\end{equation} 
The equation of motion Eq.~(5) with Eq.~(4) is used only in the slip phases, and the time intervals in the stick phases between  main shocks or between two aftershocks are theoretically calculated using Eq.~(7) or Eq.~(9). We performed a numerical simulation repeating these processes. A characteristic of our model is that a sequence of aftershocks occurs after each main shock and the difference between main shocks and aftershocks is clear.  Because $v_0$ is assumed to be infinitesimally small, the time of a sequence of aftershocks is infinitesimally small compared with the interval of successive main shocks, which is an order of $1/v_0$. 

\section{Numerical results in one dimension}
First, we show the numerical results of the modified Carlson-Langer model of ten blocks to understand the role of aftershocks. 
\begin{figure}[t]
\begin{center}
\includegraphics[height=5.2cm]{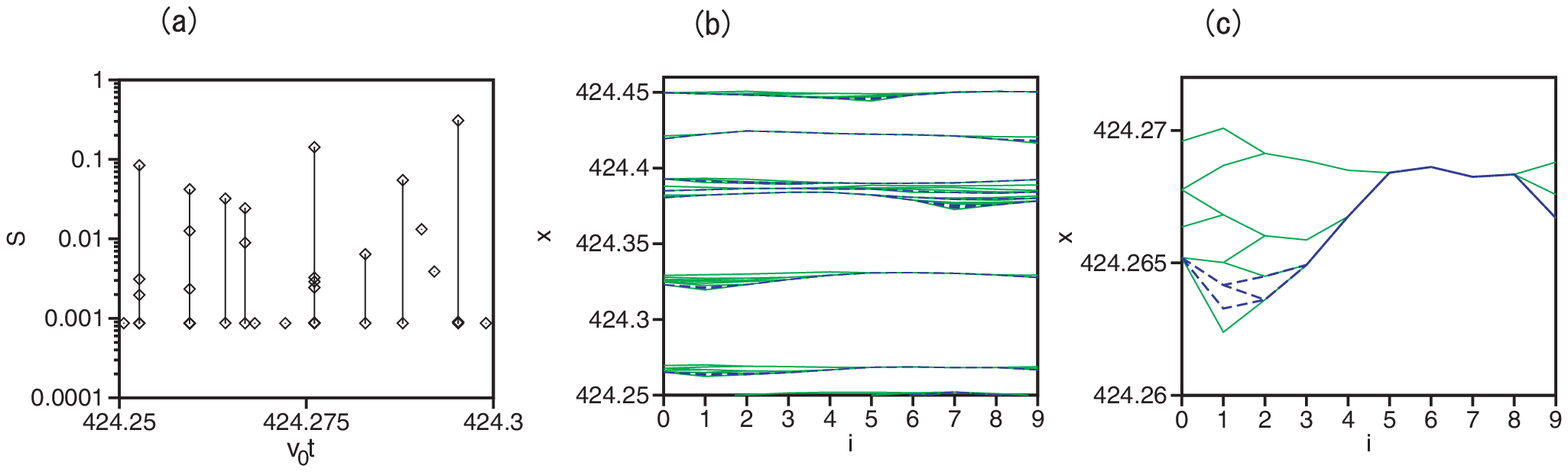}
\end{center}
\caption{(color online) (a) Time evolution of the sum $S$ of slips for ten-block systems at $\gamma=1/2$ ($\beta=2$), $\alpha=1.5,K=4$, and $\delta=K/\eta^{\beta}=1$. (b) Time evolution of the profile of $x_i$ for ten-block systems using the same parameters. The light solid lines denote the profiles of $x_i$ after main shocks and the dark dashed lines denote the profiles of $x_i$ after aftershocks. (c) Enlargement of (b) for $424.26<x<424.272$.
}
\label{f2}
\end{figure}
Figure 2(a) shows the time evolution of the sum $S=\sum s_i$ of slip distances
at $\gamma=1/2$ ($\beta=2$), $\alpha=1.5, K=4$, and $\delta=K/\eta^{\beta}=1$.  In the time scale of $O(1/v_0)$, aftershocks occur at the same time as the main shock. The vertical lines show aftershocks. The largest value of $S$ at each slip time corresponds to the main shock. A slip event without aftershocks is shown as a rhombus without vertical lines. The number of rhombi at each slip time is the sum of one (main shock) plus the number of aftershocks.  The number of aftershocks takes a various value ranging from  0 to 6. Figure 2(b) shows time evolution of the profile of $x_i$ for the ten-block system using the same parameters. The light solid lines denote the profiles of $x_i$ after the main shocks and the dark dashed lines denote the profiles of $x_i$ after the aftershocks. Figure 2(c) is the enlargement of Fig.~2(b) for $424.26<x<424.272$. After a large main shock, a dent structure is created, which is composed of delayed blocks. Similar dent structures were also found in the original Carlson-Langer model~\cite{rf:8}. Pulling forces are relatively strong in these dent regions, because $x_{i+1}-2x_i+x_{i-1}$ is positive and large. As the braking forces $f_i$ weaken, aftershocks tend to occur near these dent structures. Aftershocks make the profile of $x_i$ flat. When the profile of $x_i$ becomes sufficiently flat, a simultaneous slip or a large shock occurs next.  Through chaotic dynamics in the simultaneous slip event, the spatial fluctuation increases over time and another dent structure appears after the large-scale shock. Similar intermittent dynamics were also observed in the original Carlson-Langer model~\cite{rf:8}. In the original Carlson-Langer model, small slip evens (small-scale main shocks) frequently occurs near the dent structures. In our modified model, aftershocks play a role in these small slip events in the original Carlson Langer model.  
\begin{figure}[t]
\begin{center}
\includegraphics[height=10.cm]{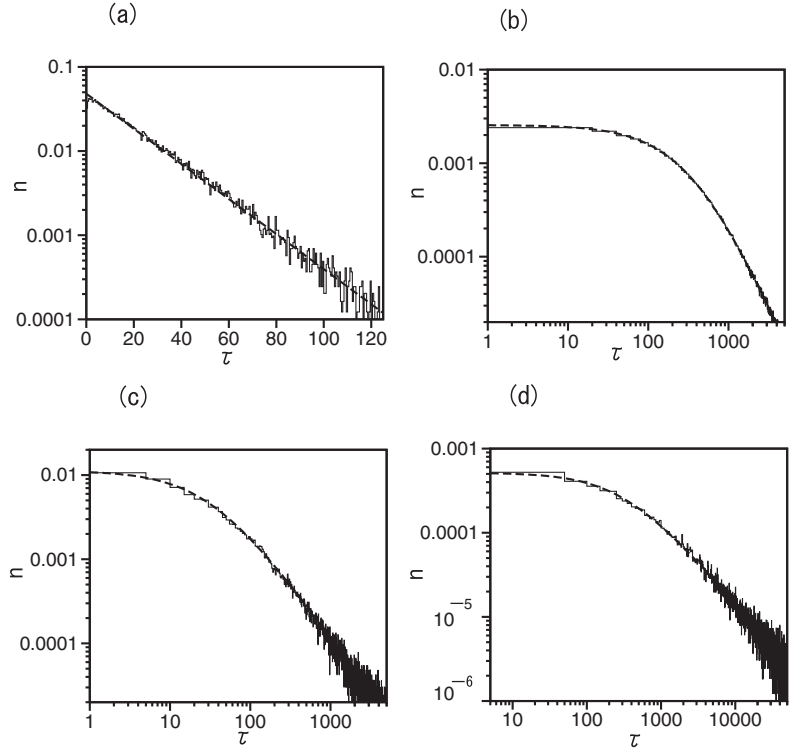}
\end{center}
\caption{(a) Semi-logarithmic plot of the probability distribution of the elapsed time $\tau$ of aftershocks at $\gamma=1,\alpha=1,K=0.5$, and $\delta=K/\eta=0.05$. (b) Double-logarithmic plot of the probability distribution of $\tau$ at $\gamma=1/2,\alpha=1,K=4$, and $\delta=K/\eta^2=1$. (c) Double-logarithmic plot of the probability distribution of $\tau$ at $\gamma=1/3,\alpha=0.5,K=0.4$, and $\delta=K/\eta^3=10000$. (d) Double-logarithmic plot of the probability distribution of $\tau$ at $\gamma=1/4,\alpha=0.5,K=0.4$, and $\delta=K/\eta^4=100000$.}
\label{f3}
\end{figure}

We investigated the statistical properties of aftershocks in a one-dimensional system of $N=1000$. We performed numerical simulations by changing parameters $\gamma, \alpha, K$ and $\delta=K/\eta^{1/\gamma}$. We present typical examples in this paper.  Figure 3(a) shows a semi-logarithmic plot of the probability distribution of the time $\tau$ when aftershocks occur after the main shocks at $\gamma=1,\alpha=1,K=0.5$, and $\delta=K/\eta=0.05$. The dashed line is $n(\tau)=0.048\exp(-0.048\tau)$, although the dashed line overlaps with the solid curve and is hardly visible. 
In this case of linear visco-elasticity $\gamma=1$, the probability distribution obeys an exponential distribution.  This implies that the simple linear model does not reproduce Omori's law of aftershocks.  Figure 3(b) shows a double-logarithmic plot of the probability distribution of $\tau$ at $\gamma=1/2,\alpha=1,K=4,$ and $\delta=K/\eta^{2}=1$. The dashed line is $n(\tau)=356/(373+\tau)^{2}$. Figure 3(c) shows a double-logarithmic plot of the probability distribution of $\tau$ at $\gamma=1/3,\alpha=0.5,K=0.4$ and $\delta=K/\eta^{3}=10000$. The dashed line is $n(\tau)=0.98/(31+\tau)^{1.3}$. Figure 3(d) shows a double-logarithmic plot of $n(\tau)$ at $\gamma=1/4,\alpha=0.5,K=0.4$, and $\delta=K/\eta^{4}=100000$. The dashed line is $n(\tau)=0.18/(310+\tau)^{1.02}$. Similarly, the probability distribution at $\gamma=3/4,\alpha=1,K=1,\delta=K/\eta^{1/\gamma}=0.1$ is also calculated and the approximate function $n(\tau)=60000/(350+\tau)^{2.8}$ is obtained. 
The probability distributions approximately obey Omori's law in the nonlinear visco-elastic models. The exponents $p$ are evaluated at $p=2.8, 2, 1.3$ and $1.02$ for $\gamma=3/4,1/2,1/3$ and $1/4$, respectively. The exponent $p$ is interpreted as $p=\infty$ for the linear case of $\gamma=1$. These results suggest that the exponent $p$ decreases with $\gamma$. For $\gamma=1/4$, the exponent $p$ is close to the exponent $p=1$ of original Omori's law.  

\begin{figure}[t]
\begin{center}
\includegraphics[height=5.cm]{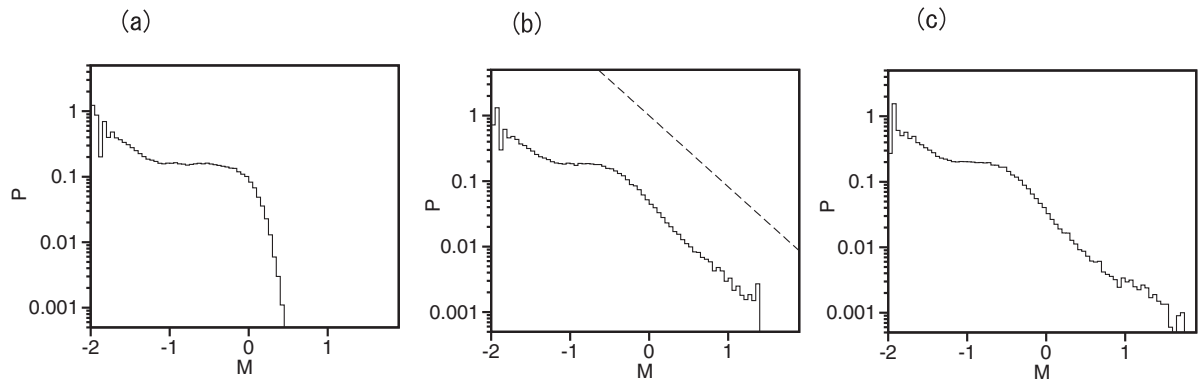}
\end{center}
\caption{Probability distribution of the magnitude at (a) $\alpha=0.3$, (b) 0.4, and (c) 0.5 for $\gamma=1/4,K=0.4$, and $\delta=K/\eta^{4}=100000$. The dashed line in Fig.4(b) denotes a curve of $P(M)\propto 10^{-1.2M}$.}
\label{f4}
\end{figure}
Figures 4 shows the probability distribution $P(M)$ of the magnitude at (a) $\alpha=0.3$, (b) 0.4, and (c) 0.5 for $\gamma=1/4,K=0.4$, and $\delta=K/\eta^{4}=100000$. Figure 4(c)  is a magnitude distribution for the numerical simulation shown in Fig.~3(d).  In the original Carlson-Langer model, the magnitude distribution has a hump structure around a rather large value of $M$ at $\alpha>1$ and decays rapidly for large $M$ at $\alpha<1$. The magnitude distibution  exhibits a power law behavior near $\alpha=1$. Even in our model with nonlinear viscoelastisity, the magnitude distribution decays rapidly for large $M$ at small $\alpha$ as shown in Fig.~4(a) and has a long tail at large $\alpha$ as shown in Fig.4(c), although there is a shoulderlike structure near $M=-0.7$ in the magnitude distributions. The magnitude distribution exhibits a power-law-like behavior at $\alpha=0.4$ for $M>-0.5$, although the linearity is not so good as compared to the models on a percolation cluster as shown in Fig.~6(a) and Fig.~7(a). The exponent $b$ is evaluated as 1.2, which is close to the $b$ value of the Gutenberg-Richter law for earthquakes.    

\begin{figure}[t]
\begin{center}
\includegraphics[height=10.cm]{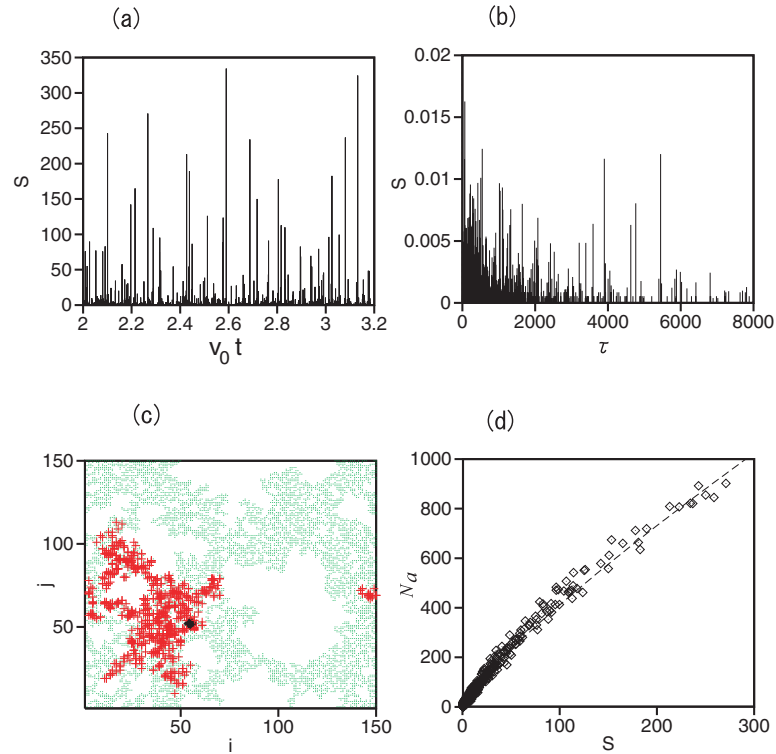}
\end{center}
\caption{(color online) (a) Time evolution of main shocks at $\gamma=1/2$,$\alpha=1.2,K=1$, and $\delta=K/\eta^2=0.05$. (b) Time evolution of aftershocks after the main shock at $v_0t=3.0812$. (c) Distribution of the epicenters of a main shock (rhombus) and the aftershocks (pluses). The critical percolation cluster is expressed using dots. 
(d) Relationship between the slip size $S$ of the main shock and the number $N_a$ of aftershocks after each main shock. The dashed line is $N_a=9S^{0.83}$}
\label{f5}
\end{figure}
\begin{figure}[t]
\begin{center}
\includegraphics[height=5.2cm]{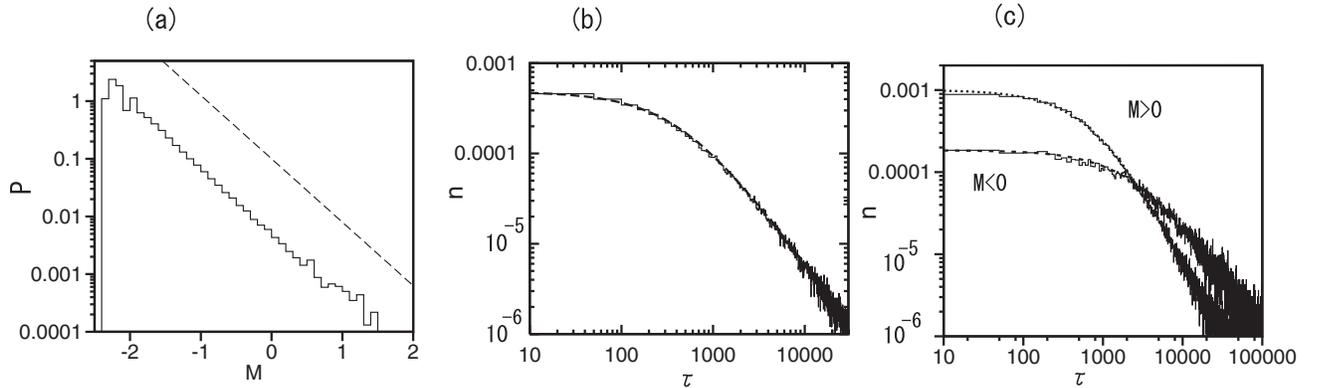}
\end{center}
\caption{(a) Probability distribution of the magnitude $M=(2/3)\log_{10} S$ at $\gamma=1/2$, $\alpha=1.2,K=1$, and $\delta=K/\eta^2=0.05$.  The dashed line denotes an exponential distribution $P(M)\propto 10^{-1.1M}$. (b) Probability distribution of $\tau$. The dashed line is $n(\tau)=13.3/(680+\tau)^{1.5}$. (c) Probability distribution of $\tau$ for $M>0$ and $M<0$. The dashed lines are $n(\tau)=7.4/(2560+\tau)^{1.35}$ and $n(\tau)=598/(916+x)^{1.95}$.}
\label{f6}
\end{figure}
\begin{figure}[t]
\begin{center}
\includegraphics[height=5.5cm]{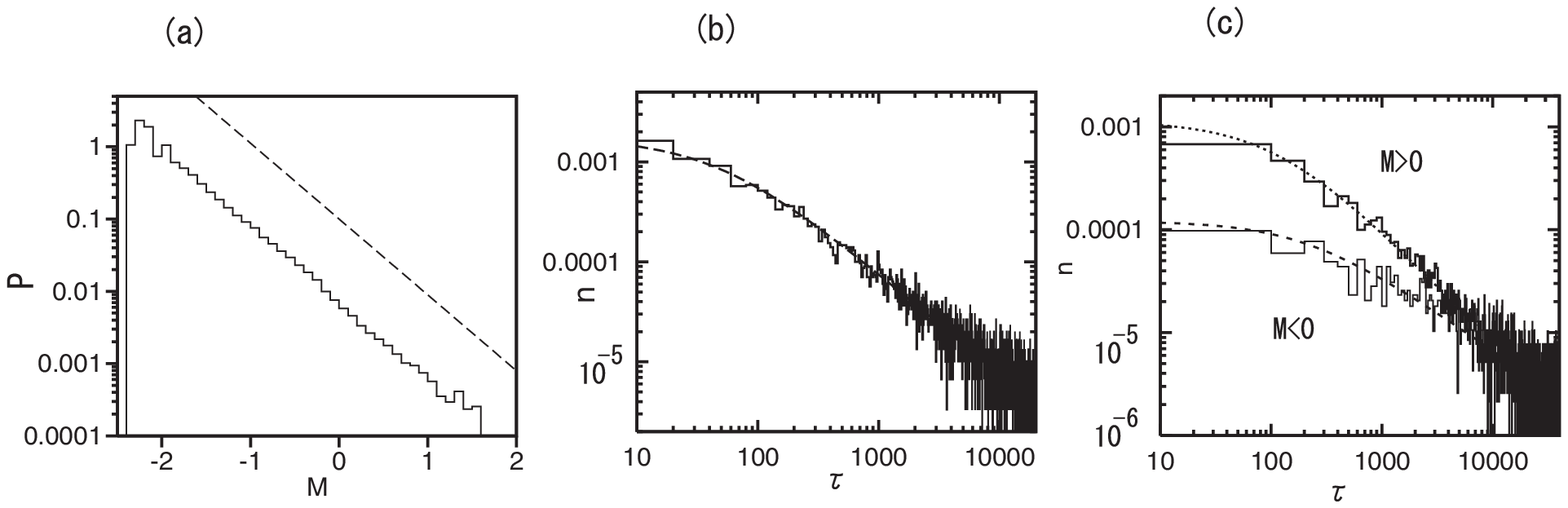}
\end{center}
\caption{(a) Probability distribution of the magnitude $M=(2/3)\log_{10} S$ at $\gamma=1/4$, $\alpha=0.95,K=0.1$, and $\delta=K/\eta^4=100000$. The dashed line denotes an exponential distribution $P(M)\propto 10^{-1.05M}$. (b) Probability distribution of $\tau$. The dashed line is $n(\tau)=0.09/(48+\tau)^{1.02}$. (c) Probability distribution of $\tau$ for $M>0$ and $M<0$. The dashed lines are $n(\tau)=0.0068/(112+\tau)^{1.1}$ and $n(\tau)=0.0068/(215+x)^{0.75}$.}
\label{f7}
\end{figure}
\section{Numerical results on a percolation cluster}
Next, we investigated the statistical properties of aftershocks on a critical percolation cluster on a square lattice of $150\times 150$.  The critical percolation cluster was constructed by the site percolation on a square lattice. Only the neighboring occupied sites are connected by springs of a spring constant $k=16$. We assumed that blocks are located only in the largest mutually-connected percolation cluster.  
In the previous paper, we performed a numerical simulation of the original Carlson-Langer model on the same critical percolation cluster~\cite{rf:8}. 
We show numerical results for the modified Carlson-Langer model on the same percolation cluster at $\gamma=1/2$, $\alpha=1.2,K=1$, and $\delta=K/\eta^2=0.05$. 
Figure 5(a) shows the time evolution of the total slip size $S$ of the main shocks.  Figure 5(b) shows the time evolution of the aftershocks after a main shock at $v_0t=3.0812$. The frequency of aftershocks decreases with time. Figure 5(c) shows epicenters of the main shock (rhombus) and aftershocks (pluses). The epicenters of aftershocks locate around the epicenter of the main shock. In real earthquakes, it is observed that epicenters of aftershocks is locally distributed around the epicenter of the main shock. Here, the epicenters are defined as the position of the block which slips first for each slip event. 
Figure 5(d) shows a relationship between the total slip size $S$ of the main shock and the number $N_a$ of aftershocks after each main shock. The number of aftershocks increases with the magnitude of the main shock.   The dashed line is $N_a=9S^{0.83}$. 

Figure 6(a) is the probability distribution of the magnitude $M=(2/3)\log_{10} S$ at $\gamma=1/2$, $\alpha=1.2,K=1$, and $\delta=K/\eta^2=0.05$. The dashed line denotes a distribution $P(M)\propto 10^{-1.1M}$. A good agreement is seen for $M<1.5$. A type of Gutenberg-Richter law is satisfied. Figure 6(b) shows the probability distribution of the elapsed time $\tau$ of the aftershocks after the main shocks. The dashed line shows $n(\tau)=13.3/(680+\tau)^{1.5}$. Omori's law of $p=1.5$ is obtained. We studied the difference of the the probability distributions of $\tau$ after the main shocks of larger magnitude $M>0$ and smaller magnitude $M<0$. Figure 6(c) shows the two probability distributions. Both distributions obey Omori's law but the exponent $p$ is 1.35 for $M>0$ and 1.95 for $M<0$. We  also performed a numerical simulation on the same percolation cluster on a square lattice of $150\times 150$ at $\gamma=1/4, \alpha=0.95,K=0.1$, and $\delta=K/\eta^4=100000$.  Figure 7(a) is the probability distribution  of the magnitude $M=(2/3)\log_{10} S$. The dashed line denotes an exponential distribution $P(M)\propto 10^{-1.05M}$. A good agreement is observed for $M<1.5$. The distribution $P(M)\propto 10^{-bM}$ with $b=1$ corresponds to the original Gutenberg-Richter law.  Figure 7(b) shows the probability distribution of $\tau$. The dashed line shows $n(\tau)=0.09/(48+\tau)^{1.02}$. Omori's law of $p=1.02$ is obtained.  The Gutenberg-Richter law and Omori's law close to the original versions are obtained at this parameter set. Figure 7(c) shows the two probability distributions of $\tau$ for $M>0$ and $M<0$. Both distributions obey Omori's law but the exponent $p$ is 1.1 for $M>0$ and 0.75 for $M<0$. We do not understand the reason but the exponent seems to depend on the magnitude of the main shock and the exponent $p$ is larger for larger $M$ in our model. We also performed a similar numerical simulation for a linear visco-elastic model of $\gamma=1$ at $\alpha=1,K=0.5$, and $\delta=K/\eta=1$. We found that the probability distribution of $\tau$ obeys an exponential function in the linear model, which is the same as the behavior of the one-dimensional model shown in Fig.~3(a). These numerical results suggest that the exponent $p$ of Omori's law does not strongly depend on the dimension. 
\section{Summary}
We have proposed a modified Carlson-Langer model which can generate aftershocks, and performed numerical simulations. The visco-elastic element works as a brake in the slip process of blocks. Owing to the damping of the braking force, the forward tensile force for blocks increases, and aftershocks are induced.   
By using nonlinear visco-elastic elements, Omori's law has been reproduced. 
On the other hand, Omori's law could not be reproduced in simple linear visco-elastic models. Both the Gutenberg-Richter law with exponent close to 1 and Omori's law with exponent close to 1 are observed in the numerical simulation of the modified Carlson-Langer model with $\gamma=1/4$ on a critical percolation cluster on a square lattice.

\end{document}